# Electrical Switching of Magnetization in Films of α-Iron with Naturally Hydroxidized Surface

N. I. Polushkin[1,2,5], A.C. Duarte[1,2], O. Conde[1,2], N. Bundaleski[3], J. P. Araujo[4],
G. N. Kakazei[4], P. Lupo[5], A.O. Adeyeye[5]

[1]*Departamento de Física, Faculdade de Ciências, Universidade de Lisboa, 1749-016 Lisboa, Portugal*
[2]*CeFEMA-Center of Physics and Engineering of Advanced Materials, 1049-001 Lisboa, Portugal*
[3]*Vinča Institute of Nuclear Sciences, University of Belgrade, P.O. Box 522 11001, Serbia*
[4]*University of Porto, Department of Physics and IFIMUP, 4169-007 Porto, Portugal*
[5]*Electrical & Computer Engineering, National University of Singapore, 17583 Singapore*

Control of the magnetization vector in ferromagnetic films and heterostructures by using electric tools instead of external magnetic fields can lead to low-power memory devices. We observe the robust changes in magnetization states of a thin (~30 nm) film of α-Fe covered by the naturally formed layer (~6 nm in thickness) of iron ohyhydroxides (FeOOH) under discharging a capacitor through the film. Strikingly, the magnetization vector is switchable by the discharge even with no any biasing field at room temperatures. In this electrically induced magnetization switching (EIMS) we reveal the key role of the FeOOH layer. We demonstrate experimental evidences that not the discharge current itself but the electric field (~10 kV/m) generated by this current is responsible for EIMS. The results reported here provide a plausible explanation of the observed phenomenon in terms of electric-field-induced weak ferromagnetism in the FeOOH layer and its coupling with the underlying α-Fe.

**Introduction.** −Manipulation of the magnetization vector in ferromagnetic materials by an electric current and/or electric field (**E**) instead of external magnetic field opens a route for integrating magnetic nanodevices into electronic circuits [1]. Magnetization switching induced by an electric current is achievable in structured materials − either via the so-called spin transfer-torque effect in lateral nanostructures [2, 3] or via the spin-orbit coupling effects in systems with lacking inversion symmetry, e.g., ferromagnetic semiconductors [4] and stacks of ultrathin layers [5]. In another kind of systems, that is, ferromagnetic metal (semi-conductor)/insulator heterostructures [6-14], the magnetization state is controllable by applying **E** and involving different mechanisms associated with modifications of the carrier concentration in magnetic semiconductors [7, 8] and of the electronic structure at the metal interfaces [9-11], or with the magnetoelectric effect in insulators with magnetic order [12-14]. Using **E** for controlling the magnetization allows for reducing the energy dissipation in the system [6]. However, the **E**-induced effects are typically so weak [15] that huge fields (up to ~$10^6$ kV/m [10, 13]), which are comparable to the threshold of electrical breakdown [16], should be applied for such a control.

We offer a paradigm of the system for which the problem of electrical aging and breakdown can be postponed. In our work we study ferromagnetic (FM) ~30-nm-thick films of α-Fe covered by the naturally formed surface nanolayer (~6 nm in thickness [17]) of iron oxyhydroxides, e.g., γ-FeOOH [18-20]. In their bulk form FeOOH polymorphs are known [20] to be antiferromagnetic (AFM), which is compatible with the temperature dependence of the coercive field in our samples. The formation of the hydroxidized layer on the surface of metals [21] results from their chemical interaction with humid atmosphere. We find that our α-Fe/FeOOH samples exhibit the robust effects of magnetization distortions [22] occurring under applying a voltage $U$=1÷100 V and discharging subsequently a capacitor through the sample [23] – with the macroscopic gap $S$ between the pads (up to $S$~5 mm). Strikingly, a change in the magnetization can reach the full reversal of its direction even with no any biasing field. Comparison of the switching thresholds measured at different temperatures ($T$) – from room temperature (RT) to liquid nitrogen one (LNT) – enables us to conclude that the in-pane electric field ($E_x$) generated by the discharge current can be responsible for the observed electrically induced magnetization switching (EIMS). This field seems to be very low ($E_x$≡$E$=$U/S$~10 kV/m) for interpreting the EIMS upon a basis of the established mechanisms. However, at least qualitatively, the occurring events can be explained in terms of weak ferromagnetism [24, 25] induced by **E** [26-27] in the AFM ion lattice of the surface FeOOH layer. Essential ingredients in such an EIMS explanation are thermal stabilization of magnetic moments in the AFM layer [28, 29] coupled with α-Fe across the interface and their out-of-plane orientation.

**Materials and methods.** – Samples were prepared onto the native oxide of Si (100) by sputtering of a pure Fe target in a vacuum chamber at base pressures $P_0$ varied between $10^{-6}$ and $10^{-8}$ mbar. After taking out a sputtered α-Fe film from the vacuum chamber to the atmosphere, the sample was naturally covered by an FeOOH layer. For sample preparation, different sputtering techniques – pulsed laser deposition (PLD) and ion-beam deposition [30] – were employed. PLD experiments, in particular, were conducted with an excimer laser LPX110i (Lambda Physik) with KrF radiation (wavelength 248 nm, pulse duration 30 ns) operating with a pulse repetition rate up to 10 Hz. A laser fluence and substrate-to-target distance used for sputtering were about 2.6 J/cm$^2$, and ~6 cm, respectively, so that the rate for Fe growth evaluated was ~0.02 nm/s.

As-prepared samples were characterized regarding their chemical and phase composition by transmission electron microscopy (TEM) and selected area electron diffraction (SAED), glancing incidence x-ray diffraction (GIXRD), and X-ray photoelectron spectroscopy (XPS). TEM/SAED data were collected with a Hitachi H-8100 electron microscope. GIXRD patterns were taken with a Siemens D5000 x-ray diffractometer in an angular range of 2θ=20÷60°. The XPS measurements were performed with a Kratos Axis Ultra spectrometer using a monochromatic Al Kα source. The most important features revealed in these studies are that the α-Fe films we sputter are nanocrystalline (about 4 nm in the averaged grain diameter) and that their outer surface is covered by a-few-nm-thick (~6 nm) layer enriched with iron oxyhydroxides (FeOOH); in greater detail see Supplemental Material – I [17].

**Experiment.** – The schematic of our experimental setup is shown in Fig. **1 (a)**. Two Al pads of a width of $W$~1.0 mm were placed onto the film surface at a pad-to-pad spacing ($S$) varying in the range from ~0.5 to 5.0 mm. The interpad gap was probed *in situ* with longitudinal magneto-optic Kerr effect (MOKE) both before and after discharging a capacitor ($C$) through the sample with a thickness of $h$~30 nm. To observe the MOKE, the probing beam from a laser diode module, operating at wavelength radiation of 670 nm with a power of 5.5 mW, was focused on the film surface between the pads. In addition, the EIMS effect was checked *ex situ* with a vibrating sample magnetometer; see Supplemental Material – II [17]. When the switch is set to the position "1", the capacitor is charged from a voltage source. Upon switching from the position "1" to "2" the capacitor is dischargeable through the sample, so that a density of the discharge current achieved is

$$j = \frac{E}{\rho} \exp(-\frac{t}{t_d}), \qquad (1)$$

where ρ is the electric resistivity, $t$ is the time elapsed from the start of a discharge, and $t_d=RC$ the discharge time. In these experiments we controlled the sheet resistance $R_s=\rho/h$. In samples produced at $P_0 \geq 5.0 \times 10^{-7}$ mbar, $R_s$ was as small as ~3 Ω/sq to be close to that quantity in bulk iron, $\rho_{iron} = 10^{-7}$ Ωm. This was achieved by removal of the high resistivity contamination (hydrocarbon) layer [31] from the surface by discharging a capacitor through the sample at a high enough voltage ($U>20$ V), so that the density of the discharge current was $j \sim E/\rho_{iron} = 10^{11}$ A/m$^2$ at $t=0$ and $E=10$ kV/m. Figure **1 (b)** shows an equivalent electric scheme of our system, where $R_\perp = \rho_{oxy}h_{oxy}/W^2$ and $R_\parallel = \rho_{oxy}S/Wh_{oxy}$ are transverse (underneath the pads) and longitudinal (along the film surface) electric resistances in the FeOOH layer with $\rho_{oxy}$ and $h_{oxy}$~6 nm [17] being its resistivity and thickness. Although $R_\perp$ is so small that does not contributes to $R_s$, $\rho_{oxy}$ can be still many orders larger than $\rho_{iron}$, so that we can assume that $R_\parallel >> R_{iron}$; in greater detail see Supplemental Material – III [17].

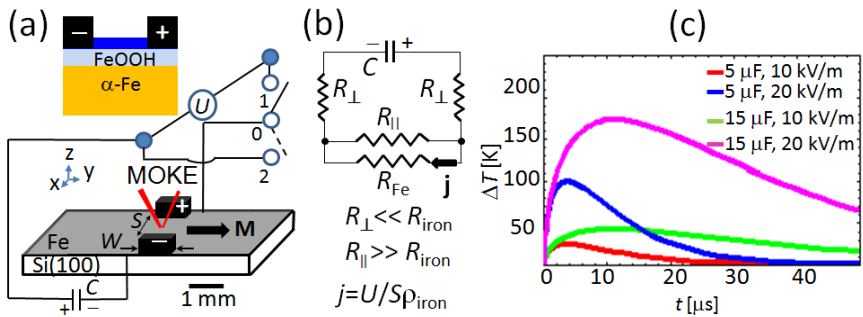

**Fig.1 (a)** Schematic of the experiment: Discharge of a capacitor through a ferromagnetic film (α-Fe) and MOKE observation *in situ*. In top we show a sketch of the system after removing preliminarily the high resistivity contamination (hydrocarbon) layer (shown in blue) underneath the pads (black boxes) by using a capacitor discharge. **(b)** Equivalent electric scheme of a capacitor discharge through an α-Fe/FeOOH bilayer; see comments through the text and in Supplemental Material – III [17]. **(c)** Calculated [Eq. (2)] temperature elevation in an Fe film on Si during a capacitor discharge as a function of time at different combinations of $E$ and $C$.

**EIMS observations**. – In order to avoid heat-assisted magnetization distortions [32], as low as possible values of $E$ and $C$ were used in our experiments, so that the current-induced temperature elevation would be small compared to the Curie temperature in α-Fe (1043K). For evaluating the current-induced temperature elevation, we follow its analytical derivation obtained in Ref. [33]

$$\Delta T(t) = 2\sqrt{\pi} E^2 \exp(-2t/t_d) \sqrt{\mu_{sub} t} / K_{sub} R_s, \quad (2)$$

where $\mu_{sub}=0.8$ cm$^2$/s and $K_{sub}=149$ W/mK are respectively the heat conductivity and diffusivity of substrate. The temperature elevation calculated according to Eq. (2) is illustrated in Fig. 1 (c). We see that in the light of this evaluation there are severe limitations for choosing $E$ and $C$.

Figure 2 presents the central result of this study, notably a capacitor-discharge-induced change of the magnetization state. In Fig. 2 (a) we show a hysteresis cycle for a sample prepared by PLD at $P_0=1.0 \times 10^{-6}$ mbar. A common feature of the samples produced at such high $P_0$ is that the discharge of a capacitor $C>1$ μF through the film at $E\sim 10$ kV/m and even with no any biasing field ($B=0$) toggles its magnetization. In Fig. 2 (a), this effect is highlighted with an arrow as a transition from the initial, $I$, (spin-up) to final, $F$, state (spin-down). Figure 2 (b) shows a sequence of discharges with probing *in situ* magnetization states. If, for instance, the system is initially in the spin-up state, applying $E>0$ and discharging subsequently the capacitor provides the change in the magnetization to spin-down. A discharge via applying $E<0$ to the spin-down state recovers the initial (spin-up) state.

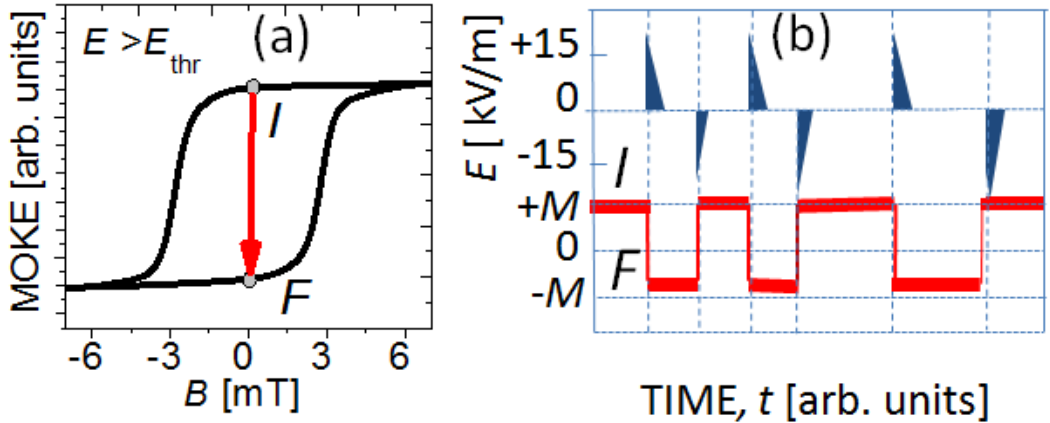

**Fig.2 (a)** MOKE hysteresis loop of a sample prepared by PLD at $P_0=10^{-6}$ mbar – with highlighting EIMS by an arrow, i.e., $I\rightarrow F$ transition, after discharging a capacitor through the sample. The state $F$ obtained by applying $E>E_{thr}$ at zero biasing field ($B=0$) is equivalent to that achievable by applying $|B|>|B_{sw}| \approx 3.0$ mT, where $B_{sw}$ is the switching field, with its subsequent switching off. **(b)** Schematic of the discharge sequence and MOKE measurements. Discharges with positive ($E>0$) and negative ($E<0$) voltages provide respectively magnetization reversals from $+M$ (state $I$) nearly to $-M$ (state $F$) and vice versa ($F\rightarrow I$).

The voltage threshold, $E_{thr}$, needed for triggering EIMS also depends on $C$, $B$, polarity of the voltage ($E$), and the angle $\phi$ between **E** and initial direction of **M**. As illustrated in Fig. 3 (a), $E_{thr}$ shifts upwards with decreasing $C$, so that a discharge time, $t_d=RC$, is not sufficient for triggering EIMS at $C<0.5$ μF (at least, at $B=0$). Figure 3 (b) shows the magnetization distortion ($\Delta$**M**) induced by a capacitor discharge as a function of $E$ for different values of $B_{sw}-B$, where $B_{sw}\approx 3.0$ mT. The data shown are retrieved for the directions of **B**≠0 both against – $B_{sw}-B=1.5$ mT and 2.0 mT – and along – $B_{sw}-B=3.3$ mT – the initial direction of **M**. Strikingly, the effect is strongly asymmetric with respect to the sign of the applied voltage. At the positive voltage applied to the capacitor, we observe $\Delta$**M** up to ~90 % [22], while the effect is found to be much weaker at negative voltages. As the direction of the voltage drop changes from its transverse ($\phi=90°$) to parallel ($\phi=0°$) orientation [Fig. 3(c)], $\Delta M$ decreases to its values near zero, at least, at $B=0$. Perhaps, the voltage drop component transverse to the magnetization orientation still persists in the $\phi=0°$ geometry and thus contributes to nonzero $\Delta M$ observed at $\phi=0°$ under a large bias,

$B_{sw}-B<1.5$ mT. It is also likely that $\Delta M$, observed at negative voltages in the $\phi=90°$ geometry and for both signs of the applied voltage in the $\phi=0°$ geometry, originates from heating effects generated by the discharge current (Eq. 2).

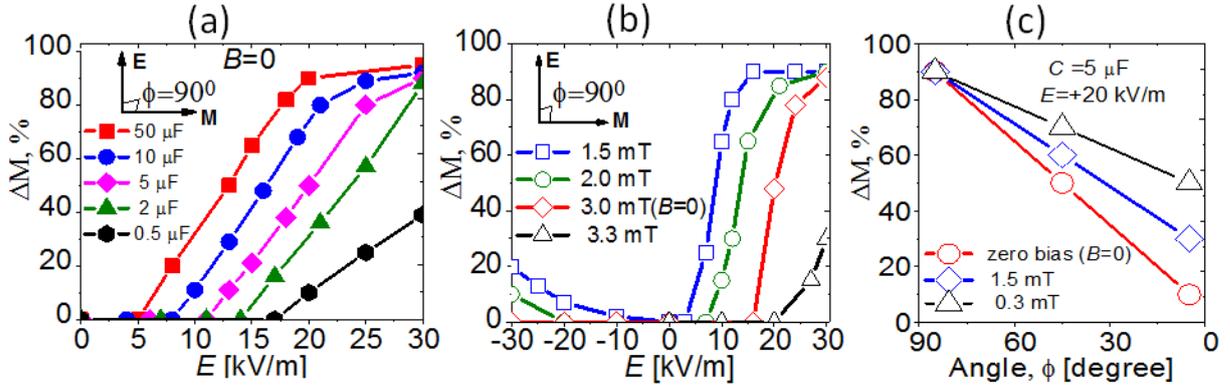

Fig. 3 (a) $\Delta M$ versus $E$ at $\phi=90°$ and zero bias ($B=0$) for different values of $C$. (b) $\Delta M$ versus $E$ for both voltage ($E$) polarities at $C=5.0$ µF and $\phi=90°$ for different values of $B_{sw}-B$. (c) $\Delta M$ as a function of the angle $\phi$ between **M** and **E** at $C=5.0$ µF and $E=+20$ kV/m for different $B_{sw}-B$.

**Role of the FeOOH layer.** – Strikingly, in α-Fe/FeOOH samples prepared at low $P_0<10^{-7}$ mbar as well as in α-Fe films covered by a layer that prevents the α-Fe surface from its hydroxidation, the capacitor discharge gives a much weaker effect on the magnetization state. We find that in both these cases the switching is achievable at much higher voltage/current thresholds, while $B$ must be close enough to the switching field, $B_{sw}$, for triggering EIMS. Figure 4 shows how the threshold current, required for triggering EIMS, depends on $B_{sw}-B$ for α-Fe/FeOOH samples prepared at $P_0=10^{-6}$ mbar (triangles) and $P_0=10^{-8}$ mbar (squares). For comparison, this dependence is also shown for a α-Fe/(8.0 nm)Au bilayer. It here is interesting that in such samples the significant magnetization distortions are observable at moderate $j$ as well (filled circles). However, we observe these changes at the applied voltage polarity (direction of **j**) opposite to that required for triggering EIMS in α-Fe/FeOOH samples. In Supplemental Material – IV

[17] we show that, in α-Fe/Au bilayers, strong effects of discharging a capacitor on the magnetization states can be attributed to the action of the Oersted field. In the context of this study, it is more important that, at the same (positive) polarity of the applied voltage, α-Fe/Au bilayers exhibit much higher current thresholds (opened circles) than those observed in α-Fe/FeOOH samples (squares and triangles). These current densities in α-Fe/Au bilayers are high enough ($j>5\times10^{11}$ A/m$^2$) to induce the heat-assisted changes in magnetization states [32, 33]. Finally, comparing the EIMS behaviors in α-Fe/FeOOH samples produced at different $P_0$, we mark that the higher $P_0$, the lower switching threshold. This fact can be explained by our microstructural studies [17] showing that in low-$P_0$ samples the amount of iron oxyhydroxides on the α-Fe surface is substantially smaller than that in high-$P_0$ samples. Therefore, we can conclude that there is the key role of the FeOOH layer in the observed EIMS.

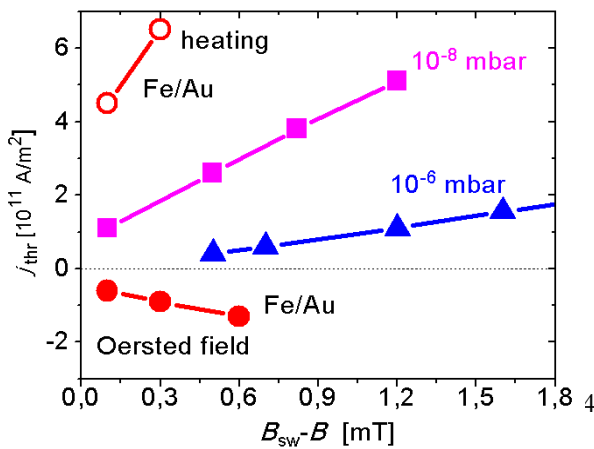

Fig.4 Switching threshold ($j_{thr}$) versus a biasing field ($B_{sw}-B$) for α-Fe/FeOOH samples produced at $P_0=10^{-6}$ mbar and $P_0=10^{-8}$ mbar and for an α-Fe/(8.0nm)Au bilayer. These data allow us to conclude about the key role of the FeOOH layer in the observed EIMS.

**Voltage/current dichotomy**. – The question of central importance is whether a discharge current or voltage applied to the gap between pads is responsible for the observed effect? To clarify this dichotomy, we measured ρ at different $T$ and compared the voltage thresholds needed for triggering EIMS with lowering $T$ from RT ($T$=295 K) to LNT ($T$=78 K). As seen from Fig. **5 (a)**, the electric properties of our samples are compatible with those in bulk of α-Fe [**34**], so that ρ(LNT) decreases by a factor of ≈10 by comparison to ρ(RT). Therefore, one can expect that the EIMS voltage threshold

$$E_{thr} = j_{thr}\,\rho(T) \qquad (3)$$

should be respectively lowered, if the discharge current ($j$) is responsible for the effect. For instance, an electric current flowing through a conductor film of a thickness of $h$ generates the magnetic field (Oersted field) whose in-plane component is $-\mu_0 j z$, where $-h/2 < z < h/2$ and $\mu_0 = 4\pi \times 10^{-7}$ Tm/A is the vacuum permeability. Therefore, the film magnetization is switchable at $j_{thr} \sim 2B_{sw}/\mu_0 h$, provided that the magnetization in its absolute value is strongly nonuniform along the film normal [**17**]. One should also take into account that, with lowering $T$, $j_{thr}$ can be altered because of shortening $t_d$ and changing the magnetization reversal features [Fig. **5(b)**].

The low-temperature data were taken by discharging a capacitor through the sample placed into a cryostat with liquid nitrogen and probing subsequently EIMS *ex situ*. Figure **5 (c)** shows the measured EIMS voltage threshold, $E_{thr}$, versus $C$ at RT and LNT. As seen, the EIMS thresholds measured at LNT are much higher (bars) than those expected under the hypothesis that the discharge current is responsible for EIMS (dashed line). The expected thresholds are obtained by multiplying $E_{thr}$'s measured at RT by a factor of $\rho_{LNT}/\rho_{RT}$ [Fig. **5 (a)**] – with no taking into account that values of $t_d$ and $B_{sw}$ are different at RT and LNT. However, from the behaviors shown in Fig. **5 (a-b)**, we see that the changes in $t_d$ and $B_{sw}$ occurring with lowering $T$ to LNT are small compared to the difference between the voltage thresholds expected at LNT and found experimentally. Moreover, these changes in $t_d$ and $B_{sw}$ must lead to shifting $E_{thr}$ in opposite directions. Due to superposition of these two effects, $E_{thr}$'s measured at LNT are close to those measured at RT (squares).

Therefore, the fact that the EIMS threshold does not depend on $T$ is indicative of the responsibility for EIMS of not the discharge current itself but the applied voltage and thus in-plane component of the electric field, $E_x \equiv E = U/S$. We admit, however, the role of the discharge current $j_x$ that generates $E_x = j_x \rho_{iron}$ which is uniform across the metallic iron. At the interface with the FeOOH layer, $E_x$ is retained by the electrodynamic boundary condition but it quickly decays ($\propto z^{-2}$) out of the layer of metallic iron [shown schematically in Fig. **5 (d)**]. In the absence of $j_x$ (blocking contacts), when a sample is additionally covered by a perfect insulator ($\rho \to \infty$), then $E_x \to 0$ near the metallic surface; in greater detail see Supplemental Material – V [**17**]. This is compatible with the well-known statement of the electrostatics [**35**].

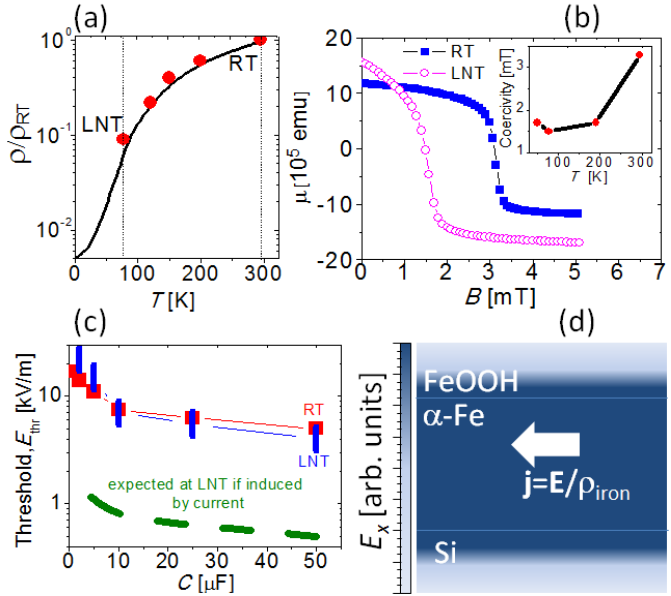

**Fig.5 (a)** The electric resistivity of bulk iron as a function of ambient temperature, $T$, taken from Ref. [**34**] (solid curve). Electric resistivity of our samples measured in a temperature range between RT and LNT is shown in the same plot by circles. **(b)** Magnetic-field-induced magnetization reversal observed at RT and LNT. The inset shows the coercivity as a function of $T$. The anomalous temperature dependence, i.e., decrease of the coercivity with lowering $T$, indicates formation of the AFM layer on top of the α-Fe. **(c)** The voltage threshold as a function of $C$ measured at RT and LNT. $E_{thr}$ obtained at RT and LNT are close each to other and strongly differ from the values expected under a hypothesis that the discharge current is responsible for EIMS.. **(d)** Schematic of the transverse (along the $z$ axis) electric-field ($E_x$) distribution generated by a discharge current flowing through the metallic iron (α-Fe).

**Hypothesis for EIMS origin**.– We believe that the events described here are interpretable in terms of **E**-induced weak ferromagnetism [24-27] occurring in the FeOOH layer that contains mostly AFM oxyhydroxide(s), notably γ–, β–, and/or hp–FeOOH [17] with Néel temperatures of 77K, 290K, and 470K, respectively [19]. Their atomic structures are based on the double chains formed by two edging sharing $FeO_3(OH)_3$ octaedra [20] with the easy AFM direction along the **c** axis [**Fig. 6 (a)**] [36]. As the coercive field decreases with lowering *T* from RT to LNT [inset of Fig. **5 (b)**]**,** we can suggest that in our samples the **c** axis mostly coincides with the direction normal to the film plane. The observed temperature dependence of the coercivity contrasts to that typically observed in FM samples (without an AFM overlayer or even with it having in-plane spin orientations) where the coercivity increases at low temperatures [28]. In our FM/AFM samples, thermal fluctuations of the magnetic moments are suppressed at low enough *T*, so that the FM spins are pinned less effectively by the out-of-plane AFM spins. In our case, the enhanced coercivity occurs just at high *T* when because of thermal fluctuations the in-plane spin projections can be large during a long enough time within the precession period of the FM spins.

In the FeOOH layer, $Fe^{3+}$ cations located at the octahedral centers are aligned antiferromagnetically by the virtue of their indirect exchange via $O^{2-}$ or $(OH)^-$ anions located at the octahedral corners. It was previously pointed out in Refs. [25-27] that, in the AFM systems with high enough crystal symmetry, the net Dzyaloshinskii vector can be equal to zero because of canceling the contributions of the superexchange pathways with opposite signs. In crystals with the certain symmetry requirements [27], an applied **E** breaks out the local inversion symmetry by inducing the ionic displacements and providing thus a nonzero Dzyaloshinskii vector, i.e., a macroscopic magnetization in the AFM component by canting magnetic moments of Fe ions, as illustrated in Figs. **6 (b-c)**.

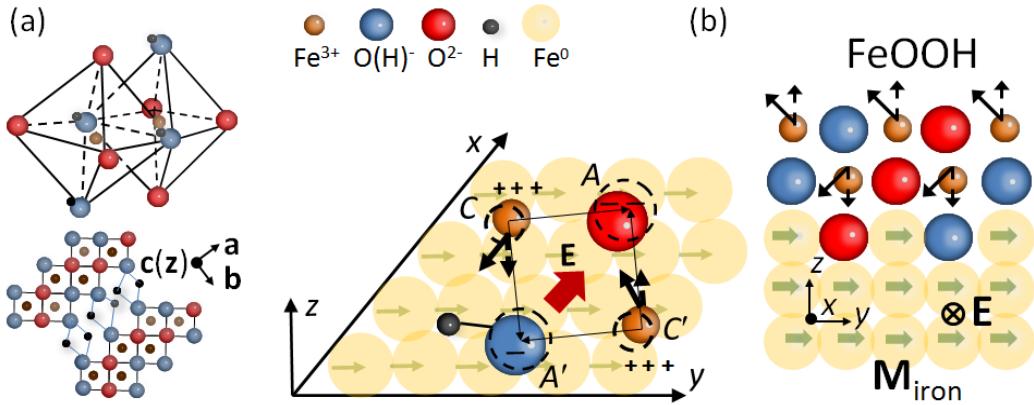

**Fig. 6 (a)** Two edging sharing $FeO_3(OH)_3$ octaedra that form the double chains in FeOOH polymorphs. Bottom: Two adjacent double chains in γ-FeOOH (unit cell parameters are *a*=0.387 nm, *b*=1.25 nm, and *c*=0.307 nm). **(b)** Proposed hypothesis for the EIMS origin: Schematic view of the atomic structure at the SNL/α-Fe interface from top (left) and from side (right). An electric field **E** ($E_x$) induces polar displacements of $Fe^{3+}$ and $O^{2-}$ ($OH^-$) ions to give rise to a nonzero Dzyaloshinskii vector and thus to a macroscopic magnetization in the FeOOH layer by canting the magnetic moments of $Fe^{3+}$ ions. This results from violating **R**$_{CA}$×**R**$_{C'A}$ = − **R**$_{CA'}$×**R**$_{C'A'}$, where **R** are the polar vectors connecting neighboring $Fe^{3+}$ cations to $(OH)^-$ and $O^{2-}$ anions. Due to interfacial exchange coupling, the $E_x$-induced magnetization in SNL is able to switch magnetic moments in the α-Fe (**M**$_{iron}$).

Because of the proximity to the FM layer with a high Curie point [28, 29], thermal fluctuations of magnetic moments in the AFM layer can be suppressed in such an extent that $E_x$ induces a magnetization $M_{oxy} > B_{sw}/\mu_0 J$ (*J* is the dimensionless interfacial exchange constant) which is able to affect the spins in the α-Fe due to interfacial coupling [12-14]. Under the out-of-plane AFM easy axis, **M**$_{oxy}$ is aligned antiparallel to the initial magnetization in the FM layer (**M**$_{iron}$). Such a configuration [Fig. **6 (c)**] is able to provide a switching of **M**$_{iron}$.

**Summary**. – We demonstrate electrically induced switching of magnetization in α-Fe films grown on Si (100) and covered by the hydroxidized surface nanolayer. We find experimental evidences that a very low electric field (~10 kV/m) applied to our bilayer samples, in which the top layer is enriched with iron oxyhydroxides like γ-FeOOH, is responsible for this switching. We believe that our finding can be promising in view of novel magnetic memory devices [37] with improved endurance and long lifetime. A density of the electric current, flowing through the conducting part of our samples and providing the electric field required for the switching, can be low enough ($\leq 10^{11}$ A/m$^2$) for reducing the undesirable effects, e.g., Joule heating (Ref. [33]). Future work will be focused on replacing the naturally formed hydroxidized layer by an artificially produced material that would have similar properties.

**ACKNOWLEDGEMENTS**

Work was supported by the Portuguese Foundation for Science and Technology (FCT) through the projects PTDC/FIS/121588/2010 and PTDC/EQUEQU/107990 /2008, and program "Ciencia 2008" (N.I.P.) We thank to C. Sá, P.A. Carvalho, and Zh. Xue for taking XPS, TEM/SAED, and a temperature dependence of the coercive field, respectively. We are grateful to A. Garcia-Garcia, S. Cardoso, and P. Strichovanek for their help in samples preparation at different stages of this work.

# Supplemental Material

## Electrical Switching of Magnetization in Films of α-Iron with Naturally Hydroxidized Surface

N. I. Polushkin[1,2,5], A.C. Duarte[1,2], O. Conde[1,2], N. Bundaleski[3], J. P. Araujo,[4] G. N. Kakazei[4], P. Lupo[5], A.O. Adeyeye[5]

[1]*Departamento de Física, Faculdade de Ciências, Universidade de Lisboa, 1749-016 Lisboa, Portugal*

[2]*CeFEMA-Center of Physics and Engineering of Advanced Materials, 1049-001 Lisboa, Portugal*

[3]*Vinča Institute of Nuclear Sciences, University of Belgrade, P.O. Box 522 11001, Serbia*

[4]*University of Porto, Department of Physics and IFIMUP, 4169-007 Porto, Portugal*

[5]*Electrical & Computer Engineering, National University of Singapore, 17583 Singapore*

## I. MICROSTRUCTURAL STUDIES

Figure **S1 (a-b)** shows a SAED pattern for an Si(100)/α-Fe/FeOOH sample prepared by PLD at $P_0=1.0\times10^{-6}$ mbar, where we mark up to four diffraction rings − with the ratios between their radiuses consistent to the inverse values of the interplanar distances in α-Fe. Figure **S1 (c-d)** illustrates GIXRD data collected for the same sample in two geometries: The x-ray incidence plane is parallel to either the (110) or (100) plane. As seen, in the GIXRD pattern taken under these configurations there are two discerned peaks. One of them arises at $2\theta\approx44.5°$ to be attributed to a (110) reflection of α-Fe. Its full width at half of maximum is about $2\Delta\theta_{0.5}\approx4.5°$, so that the averaged grain diameter in the α-Fe evaluated according to the Scherrer equation is ~4 nm. Another peak appears to be at $2\theta\approx27.5$. Among all known iron oxides and hydroxides, there are three oxyhydroxide polymorphs − γ-FeOOH (lepidocrocite), β-FeOOH (akageneite), and high-pressure (hp) FeOOH − give relatively strong reflections at this angle [**S1**]. All of these phases are known [**S1**] to be antiferromagnetic in bulk − with Néel temperatures of 77 K, 290 K , and 470 K. As the GIXRD patterns indicate the presence, at least, of the two phases, contrary to the SAED data that reveal α-Fe

only, the GIXRD data confirm the formation of the bilayer structure, i.e., the α-Fe layer covered by the surface layer which is enriched with the FeOOH polymorph(s).

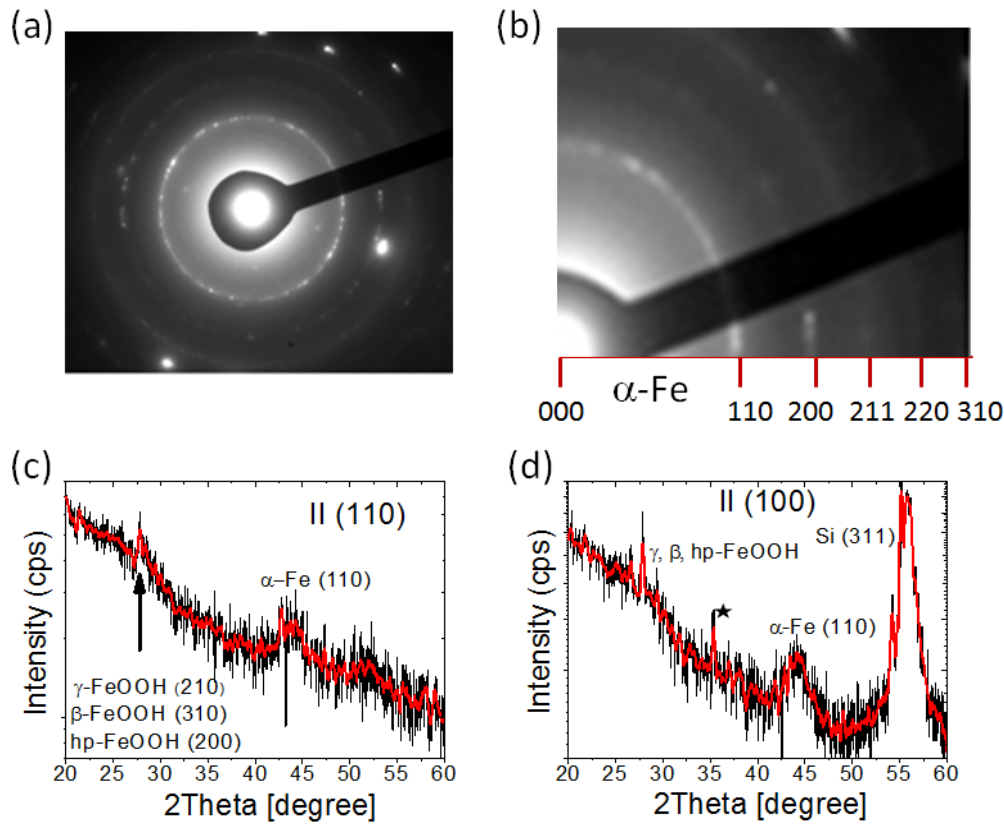

**Fig. S1. (a)** Electron diffraction pattern (SAED) taken in the transmission mode (TEM) for a sample prepared by PLD at $P_0=1.0\times10^{-6}$ mbar. **(b)** A quarter of the same SAED pattern that is shown in **(a).** There is good agreement with inverse planar distances of α-Fe. **(c-d)** GIXRD patterns taken at a 2-degree grazing angle and the two orientations of the x-ray incidence plane with respect to the sample. The two discerned peaks respectively indicate the presence of the oxyhydroxide phase(s) – γ-FeOOH, β-FeOOH, and/or hp-FeOOH – at 2θ~27.5° **[S1]** and α-Fe at 2θ~44.5°. A 'star' symbol in the pattern taken under the II (100) geometry **(d)** indicates a spurious peak.

Our conclusion about composition of the surface layer is corroborated by XPS data that allow for a high-sensitive analysis of oxides and oxyhydroxides at the film surface [S2-S4], including the determination of the thickness of the surface hydroxidized layer [S2]. Figure **S2** shows experimental XPS data (opened circles) collected in the O 1s and Fe $2p_{3/2}$ core line regions for samples produced at $P_0=10^{-6}$ mbar and $P_0=10^{-8}$ mbar (reference). Obtained spectra were calibrated by using the C 1s line which is assumed to arise in the spectrum owing to adventitious carbon adsorbate. After determining the Shirley background [S5] and employing CasaXPS software used for analyzing XPS data [S6], we have found that for the high-$P_0$ sample there are three major contributions into the O 1s line. The strongest peak is centered at a binding energy $E\approx531.6$ eV (peak 1) to be attributed to an FeOOH ($Fe^{3+}$) polymorph(s), while a lower peak arising at $E\approx530.0$ eV (peak 2) is associated with an iron oxide(s) [S2-S4]. The weakest of these three peaks centered at $E\approx532.8$ eV (peak 3) relates to the surface contamination (C-O and/or adsorbed $H_2O$).

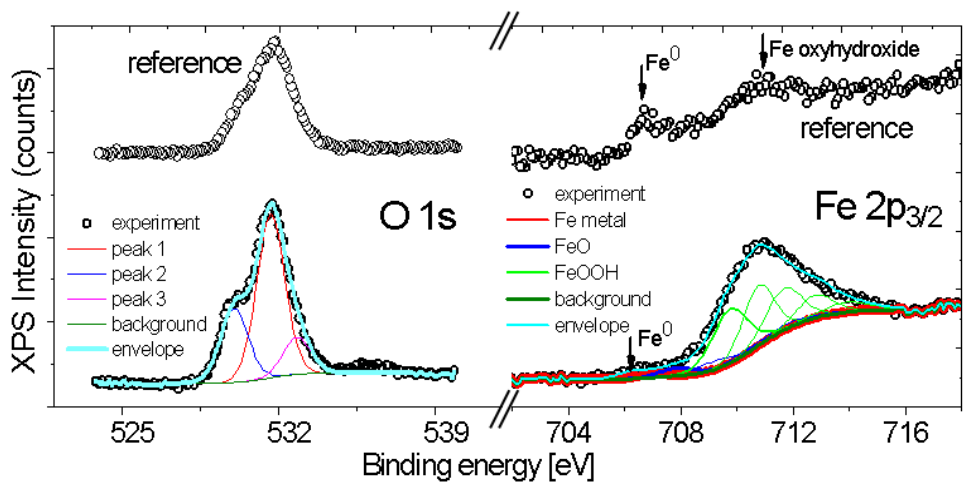

**Fig. S2**. XPS spectra in the O 1s and Fe $2p_{3/2}$ regions for $\alpha$-Fe/FeOOH samples produced at $P_0=10^{-6}$ mbar (bottom) and $P_0=10^{-8}$ mbar (top). Our analysis of the XPS data reveals a strong contribution of an FeOOH ($Fe^{3+}$) polymorph(s) into composition of the surface layer. In the low-$P_0$ sample (reference) the amount of iron oxyhydroxide/oxide phases is substantially smaller than that in the high-$P_0$ sample.

At least, for the high-$P_0$ sample, the O 1s line and its analysis well agree with the XPS data taken in the Fe 2p region. This line is very broad owing to the multiplet splitting [S7], so that each chemical bond has to be described by several peaks with well defined constraints i.e. fixed relative positions, relative intensities, and peak widths. In this approach, there are only two fitting parameters per each contribution – position and intensity of the first peak. The optimum fit for this line was obtained as a superposition of three different contributions associated to FeOOH ($Fe^{3+}$), FeO ($Fe^{2+}$) and metallic iron ($Fe^0$). The constraints for each contribution were taken from Ref. [S4]. Assuming the three compounds are uniformly distributed, composition of the surface layer can be determined as 3.9 % of metallic iron, 13.3 % of FeO and 82.8 % of FeOOH. However, this is not the case: Metallic iron is covered by a thin hydroxidized layer whose thickness can be evaluated as [S2]

$$h_{oxy} = \lambda \ln(1 + I_{oxy}/I_{iron}) \sim 6.0 \text{ nm} \quad (S1)$$

where $\lambda \sim 2.0$ nm is the attenuation electron length and $I_{oxy}/I_{iron} \approx 10$ is the ratio of peak intensities of the FeOOH phase and metallic iron. We see that, in the low-$P_0$ sample, the amount of iron oxyhydroxides is substantially smaller. Indeed, the peak intensity of metallic iron in this sample is even larger than that of $Fe^{3+}$ ions in the FeOOH phase.

A question arising in these studies is why does the amount of oxyhydroxides on top of the α-Fe depend on $P_0$? A possible explanation is that the volume inclusions of oxygen and hydrogen (or $H_2O$) play a role in formation of the surface hydroxidized layer, which results from penetration of oxygen/hydrogen atoms (molecules) into the subsurface layer from their flux that exposes the sample after opening the vacuum chamber. Newly arrived adsorbates are involved into subsurface hydroxidation via the process of Ostwald ripening [**S8**]. The higher $P_0$ during the film growth, the more oxygen/hydrogen nucleation centers are inside the Fe, and the higher rate of the (sub)surface hydroxidation after opening the vacuum chamber.

## II. EIMS TEST WITH VSM

In addition to controlling *in situ* the EIMS process with MOKE, we verified it *ex situ* with a Lake Shore vibrating sample magnetometer (VSM) system. For the latter test, we have used two identical α-Fe/FeOOH specimens **(a)** and **(b)**, with lateral sizes of $L \approx 2.5 \times 2.5$ mm$^2$. Both specimens were preliminarily magnetized along the film plane in the same direction up to saturation. Then, the specimen **(a)** was mounted in the EIMS setup [Fig. **1(a)**] for discharging a capacitor $C$=5.0 μF through it with no a biasing field ($B$=0) at $S \sim 1.5$ mm and $U$=+30 V. According to the EIMS data illustrated in Fig. **3**, these parameters were sufficient for the switching. At the next step, both specimens were mounted into the VSM system, one by one, for measuring their magnetic moments in external magnetic fields sweeping from zero to +10 mT and back (shown by arrows). We find that the changes of the magnetic moment in the cases **(a)** and **(b)** are strongly different, as shown in Fig. **S3**, including the fact that the initial magnetic moments are of different polarity and taking into account that only a part of the sample **(a)** could be repolarized because of $S<L$. This experiment provides straightforward evidence for switching of magnetization under discharging a capacitor.

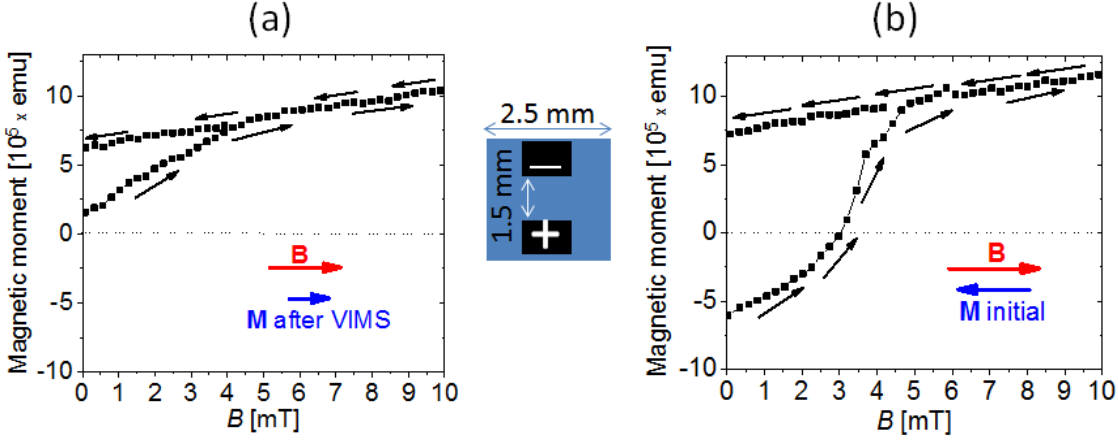

**Fig. S3**. Magnetization reversals of two identical specimens with different magnetization states prepared preliminarily for taking VSM measurements: **(a)** The magnetization of one of two specimens was switched by a capacitor discharge before taking the VSM measurement; **(b)** In another specimen, the magnetization was kept in its initial state before taking the VSM measurement.

## III. EQUIVALENT ELECTRIC SCHEME

The equivalent electric scheme of our bilayer system is shown in Fig. **1 (b)** and reproduced here (**Fig. S4**). The electric resistance of our system is

$$R = 2R_\perp + \frac{R_\| R_{iron}}{R_\| + R_{iron}}, \qquad (S2)$$

where $R_\perp = \rho_{oxy} h_{oxy}/W^2$ is the transverse (underneath the pads) resistance of the surface layer, $\rho_{oxy}$ and $h_{oxy}$ are the resistivity and thickness of the surface layer, $W$ the width of the pads, $R_{iron} = \rho_{iron} S/W h_{iron}$ the electric resistance of the metallic layer, $\rho_{iron}$ and $h_{iron}$ are the resistivity and thickness of the metallic iron, $S$ the interpad gap, and $R_\|$ is the in-plane (between the pads) resistance of the surface layer. The latter quantity can be expressed as

$$R_\| = R_{cond} R_{FeOOH}/(R_{cond} + R_{FeOOH}), \qquad (S3)$$

where $R_{cond}$ is the overall resistance of the conductive paths in the hydroxidized layer and $R_{FeOOH}$ the resistance of the oxyhydroxide phase. Clearly, as FeOOH is an insulator, $R_{FeOOH} >> R_{cond}$, so that $R_{\parallel} \approx R_{cond}$. As our measurements of the sheet resistance $R_s = \rho/h$ of our system reveal, this quantity is very close to that of Fe everywhere within a temperature between RT and LNT, so that

$$\rho(T) \approx \rho_{iron}(T), \qquad (S4)$$

which is the same that

$$R(T) \approx R_{iron}(T) \qquad (S5)$$

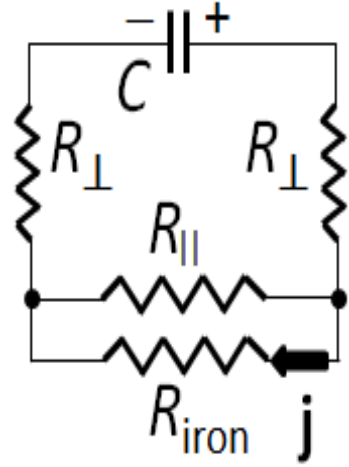

**Fig. S4.** Equivalent electric scheme of our bilayer structure: An α-Fe film (metallic iron) covered by the hydroxidized surface layer.

Therefore, taking into account Eqs. (1)–(2) and that the top (hydroxidized) layer is much thinner than that of metallic iron, we get

$$R_{\parallel}(T) \approx R_{cond}(T) >> R_{iron}(T), \text{ while } R_{\perp}(T) << R_{iron}(T) \qquad (S6)$$

## IV. MAGNETIZATION DISTORTIONS IN SAMPLES PREPARED AT LOW PRESSURES AS WELL AS WITH CAPPING

We have compared the EIMS behaviors in the $\alpha$-Fe (Fe) samples prepared at different base pressures ($P_0$) and produced with both natural (hydroxidized layer) and artificial (Au layer) capping on the Fe surface. Figure **S5** shows the magnetization distortion, $\Delta \mathbf{M}$, as function of $E$ in Si/Fe/FeOOH samples prepared at low ($P_0 = 10^{-8}$ mbar) (circles) and high ($P_0 = 10^{-6}$ mbar) (bars) base pressures. In order to highlight the effects of capping, in the same plot we show $\Delta \mathbf{M}(E)$ dependences for Si/Fe/(8nm)Au/ (rhombs) and Si/(8 nm)Au/Fe/FeOOH (squares) samples prepared at $P_0 = 10^{-8}$ mbar.

Strikingly, the EIMS threshold observed for the high-$P_0$ sample is significantly lower than that for the low-$P_0$ sample. Moreover, in the low-$P_0$ sample the EIMS is observable only at biasing fields $B$ close to the switching field, $B_{sw} \approx 3.0$ mT. As revealed by our XPS analysis, the amount of the ohyhydroxide phase becomes substantially smaller in samples sputtered at low $P_0 < 10^{-7}$ mbar.

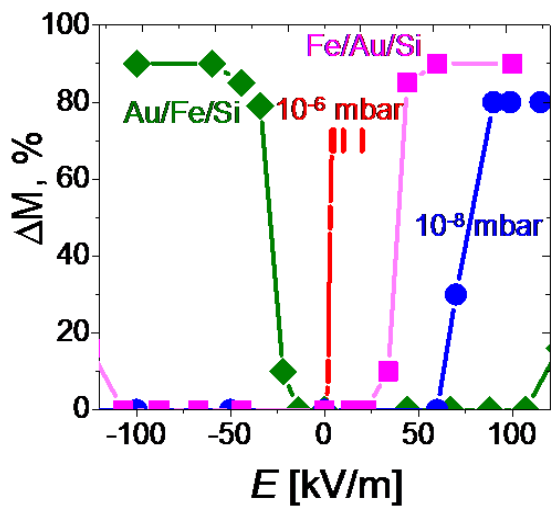

**Fig. S5.** Magnetization distortion, $\Delta M$, versus $E$ in Fe/FeOOH samples prepared on Si (100) at $P_0=10^{-6}$ mbar (bars), $P_0=10^{-8}$ mbar (circles), and with an 8-nm-thick Au layer on top (rhombs), that is, instead of FeOOH, and bottom (squares) of the α-Fe (Fe) layer. In the high-$P_0$ sample, the threshold for $\Delta M$ induced by a capacitor discharge is significantly lower than in low-$P_0$ sample as well as in those covered by Au instead of FeOOH. A biasing field used for plotting these dependences was $B_{sw}-B=0.1$ mT for the low-$P_0$ samples (Si/Fe/FeOOH, Si/Fe/Au, and Si/Au/Fe/FeOOH) and $B_{sw}-B=0.5$ mT for the high-$P_0$ sample (Si/Fe/FeOOH).

It is also interesting that in the Si/Fe/Au sample the magnetization distortion occurs at the voltage polarity opposite to that providing the EIMS effect in the α-Fe covered with FeOOH. However, contrary to Si/Fe/Au, the Si/Au/Fe/FeOOH sample displays EIMS at the same voltage polarity that gives EIMS in the Si/α-Fe/FeOOH samples. Our simulations with CST Microwave Studio software [S9] indicate that a realistic discharge provides the Oersted-field asymmetry which is able to affect the magnetization provided that the contribution of the Oersted field is added to that of a biasing field $B$ close enough to $B_{sw}$. The results of these simulations are shown in Figs. **S6** and **S7.** We see that the effect of the Oersted field increases in bilayers with a low-resistive cap (Fig. **S7**). In Au/Fe bilayers, however, the EIMS threshold is much higher than that in high-$P_0$ samples covered by a relatively high resistive hydroxidized layer (Fig. **S5**). Therefore, we were able to discriminate the effect of the Oersted field from the EIMS effect reported here.

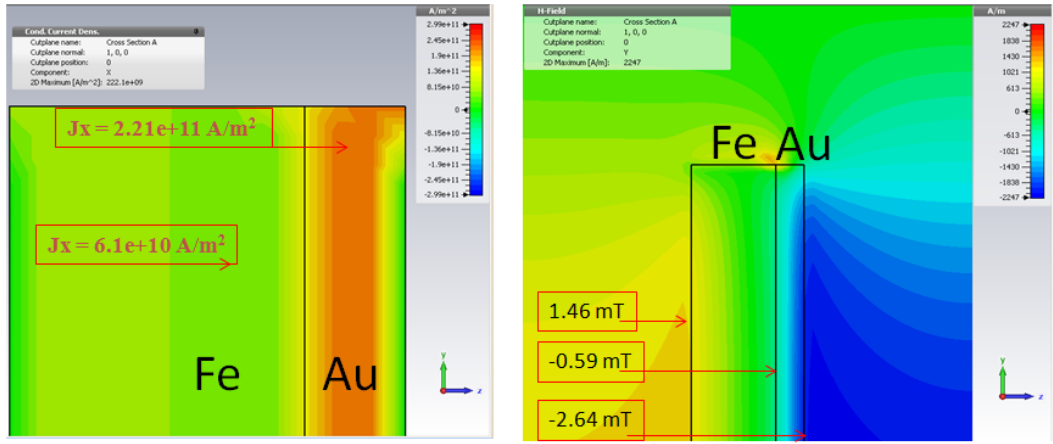

**Fig. S6.** Simulated distributions of the electric current (left) flowing in an (25 nm)Fe/(8 nm)Au bilayer along the *x* axis and generated *y*-component of the Oersted field across the bilayer structure (right). A difference between the Oersted fields at the edges of the Fe layer can be sufficient (0.87 mT) for a distortion of the magnetization by the Oersted field added to a biasing field $B_{sw}-B \sim 2.0$ mT ($B_{sw}=3.0$ mT)

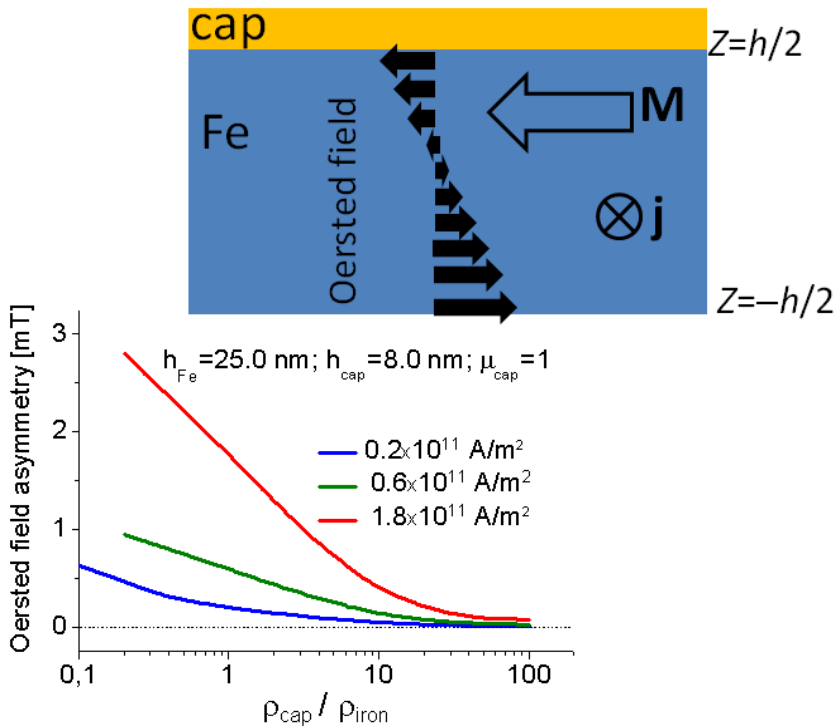

**Fig. S7.** Simulated difference between the Oersted field strengths at the edges of an Fe layer, $z=-h/2$ and $z=h/2$, as a function of the electric resistivity of the cap layer, $\rho_{cap}$. The dependence is calculated for a 25-nm-thick Fe layer with a 8.0-nm-thick cap for three values of the maximal current density inside the Fe layer. In top, the Oersted field distribution across the Fe layer in a bilayer structure is shown schematically. If both $\rho_{cap}$ and the cap magnetization could be equal to that in Fe, the asymmetry of the Oersted field would be equal to zero because of cancelling the Oersted-field torques each with other across the Fe. With a nonmagnetic cap such an Au having a resistivity much lower than that in Fe, the Oersted field asymmetry might be high enough to affect the magnetization in the Fe at realistic $j$ (**Fig.S6 left**). Experimentally, we observe this effect in Au/Fe bilayers on Si at the voltage polarity opposite to that needed for triggering the effect reported here. Also, by contrast, biasing fields for the effect of the Oersted field should be very close to the switching field. We see that the Oersted field asymmetry quickly decays with increasing $\rho_{cap}$, so that the Oersted field with the hydroxidized capping having a high $\rho_{cap}$ is not able to induce switching of magnetization in Fe.

## V. CALCULATION OF ELECTRIC FIELD WITH BLOCKING CONTACTS

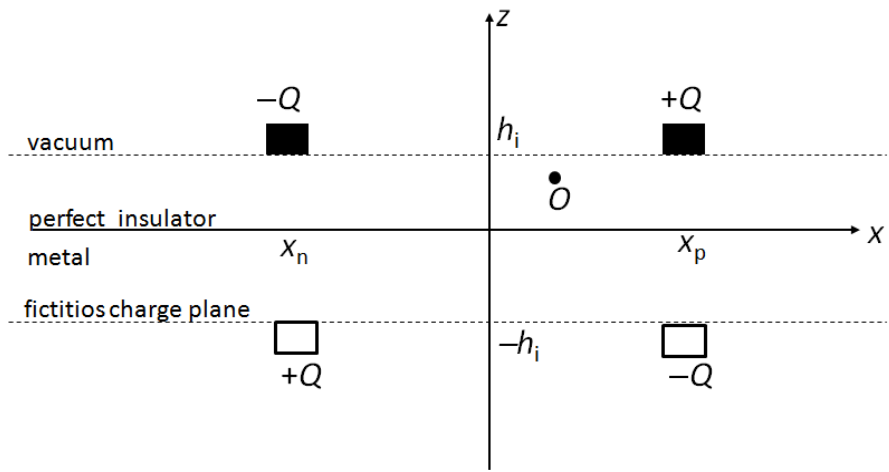

**Fig. S8**. Geometry of the system with blocking contacts

Figure **S8** shows the geometry of the system with blocking contacts, i.e., electric pads lying on the layer of a perfect insulator with a thickness of $h_i$ that covers a metal film having a thickness $h \gg h_i$. Within an approximation where the distance between the pads is much compared to the width of the pads, $W$, the pads can be replaced by point charges, $+Q$ and $-Q$, and the electrostatic potential in the observation point $O$ is given by [**S10**]

$$\varphi = Q\left(\frac{1}{r_p} - \frac{1}{r'_p} - \frac{1}{r_n} + \frac{1}{r'_n}\right), \tag{S7}$$

where $r_p$, $r'_p$, $r_n$, and $r'_n$ are the distances between $O$ and real and fictitious charges. According to the notations of Fig. S8, these distances are as follows

$$r_p = \sqrt{(x_p - x)^2 + (h_i - z)^2} \quad r'_p = \sqrt{(x_p - x)^2 + (h_i + z)^2}$$
$$r_n = \sqrt{(x_n - x)^2 + (h_i - z)^2} \quad r'_n = \sqrt{(x_n - x)^2 + (h_i + z)^2} \quad , \quad (S8)$$

where $x$ and $z$ are the coordinates of the observation point $O$ (any dependence on $y$ is omitted). Then, the longitudinal component of electric field can be written as

$$E_x \equiv -\frac{\partial \varphi}{\partial x} = Qx\left[\frac{1}{[(x-x_p)^2 + (z-h_i)^2]^{3/2}} - \frac{1}{[(x-x_p)^2 + (z+h_i)^2]^{3/2}} + \frac{1}{[(x-x_n)^2 + (z+h_i)^2]^{3/2}} - \frac{1}{[(x-x_n)^2 + (z-h_i)^2]^{3/2}}\right]. \quad (S9)$$

It is seen from Eq. (S9) that $E_x \to 0$, if $|x| \gg |z| \sim h_i$, which is compatible with the geometry of our system. This fact is just the well-known statement of the electrostatics [S10] that the tangential component of the electric field near a metallic surface must be equal to zero.